\documentclass[a4paper,11pt]{article}
\usepackage{graphicx}
\usepackage{jheppub} % for details on the use of the package, please
\usepackage{amsmath}
\usepackage[T1]{fontenc} % if needed

\title{Conformal Transformations, Rotating String and Effects of angular velocity on Accelerating Quark-Antiquark pair in $AdS_3$}

\author[a,1]{J. Sadeghi,}
\author[a,2]{F. Razavi,}
\affiliation[a]{Sciences Faculty, Department of Physics, University of  Mazandaran, Iran}
\emailAdd{pouriya@ipm.ir}
\emailAdd{f.razavi@stu.umz.ac.ir}

\abstract{In order to study quark and anti-quark interaction, one should consider all effects of the medium in motion of the pair. Because in QGP the pair is not produced at rest. So the velocity of the pair, has some effects on its interactions that should be taken into account. In this paper we apply some conformal transformations to a rotating string dual to a rotating heavy quark in $AdS_3$ which construct an accelerating string dual to an accelerating quark and anti-quark pair. So, when pair has angular velocity, it can be compared with the situation which it has just constant velocity. Then we can study effects of angular velocity on the accelerating quark and anti-quark which are constructed by performing special conformal transformations, conformal SO(2,2) transformation and particular $SL(2,R)_L$ and $SL(2,R)_R$ transformation. The accelerating quark and anti-quark show different behavior with increasing in angular velocity. With useful numerical solutions we show that quark and anti-quark can deccelerate to achieve each other or accelerate to get away from each other. We will see different behaviors of the pair in each particular transformations. There are various behavior here, such as: permanent, increase or decrese in acceleration and uncertain behaviors.}

\begin{document} 
\maketitle
\flushbottom
\section{Introduction}
AdS/CFT is a correspondance \cite{JM, GKP, EW, AGM} between a string theory in AdS space and a conformal field theory in physical space-time. It leads to an analytic semi-classical model for strongly coupled QCD. The moving open string with constant velocity solutions have been constructed by using Nambu-Goto action in $AdS_5$ black hole backgrounds \cite{HKY, SG}. In order to study the dynamics of quark moving in strongly coupled $N = 4$ SYM thermal plasma two cases are investigated. First case is The infinitely massive quark as the open string end at the boundary of the $AdS_5$-Schwarzschild space-time \cite{SG}. And second is the finitely massive quark as the open string end at the D7-brane \cite{KK} in the $AdS_5$-Schwarzschild space-time \cite{HKY}.\\
Mikhailov has introduced an analytic generic solution for the open string dual to a single infinitely massive quark moving on an arbitrary timelike trajectory in the $N = 4$ SYM theory. And he extracted a rate of the energy loss which agrees with the Lienard formula by using the Nambu-Goto action in the Poincare $AdS_5$ spacetime \cite{AM}. Based on the extension of this generic solution to the finite quark mass case it has been shown that an event horizon appears on the worldsheet whenever the single quark accelerates in any fashion \cite{CG}. By applying AdS/CFT correspondence, the dynamics of a composite quark or dressed quark \cite{CGG, MCGG} in strongly-coupled large-Nc $N = 4$ SYM has been studied. Bo-Wen Xiao has used the Nambu-Goto action in the Poincare $AdS_5$ spacetime and found an accelerating open string solution dual to a heavy quark-antiquark pair uniformly accelerated in opposite directions \cite{BX}. In that case the string configuration has expressed as
\begin{equation}
x = \pm \sqrt{t^2 + b^2 - z^2}.
\label{ex}\end{equation}
in this expression, the plus and minus sign of $x$ represents position of the quark and antiquark respectively and  $\frac{1}{b}$ is the acceleration of the quark and anti-quark pair \cite{BX, SRG}. Here the infinitely massive quark and antiquark are located on the hyperbolic trajectories $x = \sqrt{t^2 + b^2}$ at the AdS boundary z = 0 such that the plus and minus sign of (\ref{ex}) represents the right and left half of the accelerating string. The quark and antiquark first approach to each other in decelerating, and stop to return back in accelerating away from each other with proper acceleration $\frac{1}{b}$ \cite{SRG}. The Nambu-Goto action has been analysed in Rindler spacetime to construct the similar accelerating open string solution \cite{PPZ}. The energy loss is related with the appearance of the worldsheet horizon via the open string which is moving in $AdS_5$-Schwarzschild spacetime \cite{SSG,JCT, BMX} and the worldsheet Hawking radiation generates the stochastic motion of the quark \cite{BHR,ST}. \\
In \cite{BX} the accelerating string solution which is related to uniformly accelerating quark and anti-quark, the resulting stochastic fluctuations of the quark trajectory have been determined \cite{CCG}. The stochastic motion of the string is due to interactions with other Hawking quanta in de Sitter space has been studied in \cite{WPJW}. Also for the case where the flux tube is sourced by a uniformly accelerated quark it is determined that the dual string embedding known heretofore terminates unphysically at the worldsheet horizon \cite{JAP}. The generic string solution has been directly substituted into the string equation of motion and confirmed to solve it \cite{GP}. There are several researches about thermal effects of the worldsheet horizon on the accelerating string associated with Unruh temperature \cite{CCG, CP, HKK}. By applying the string solution \cite{AM} and suitable coordinate, the accelerating string solution dual to a single accelerating quark in the global $AdS_5$ spacetime has been constructed \cite{HS, FGP}.
The entanglement of the general quantum Einstein-Podolsky-Rosen (EPR) pair are connected through the interior via a wormhole, or Einstein-Rosen bridge \cite{MS}. There have been various investigations where the quark-antiquark pair is concretely regarded as a color singlet EPR pair in the $N = 4$ SYM theory and the related entanglement is encoded in a nontraversable wormhole on the worldsheet of the flux tube connecting the pair \cite{JK, JS, CGP, JKR}.In \cite{WPJ} the string solution dual to a quark and anti-quark pair with constant acceleration in de Sitter space has been constructed and shown that quark and anti-quark pair can be interpreted as a EPR pair or Hawking pair emanating from dS horizon. These results are related to the existence of horizon on the worldsheet
of the accelerating open string dual to a uniformly accelerating quark-antiquark pair. The entanglement entropy of the quark-antiquark pair has been studied in \cite{JK, LM, VHGS}. And the relevence of the entanglement entropy and the string surface describing gluon scattering in position space has been investigated \cite{SS}.\\
In \cite{SRY} the two string solutions associated with One-cusp Wilson loop \cite{MK} has been reconstructed which are dual to an accelerating quark-antiquark pair \cite{BX}. The $SL(2,R)L \times SL(2,R)R$ isometry group of the $AdS_3$ space-time has been studied in \cite{JMS,BHT}.By performing the conformal SO(2,4) transformation, one-cusp Wilson loop solution has been studied, which provides chances to study the planar four-gluon scattering amplitude by using four-cusp Wilson loop solution in the T-dual AdS space-time \cite{LAM}. An accelerating open string solution dual to a heavy accelerating quark and anti-quark pair in opposite direction has been constructed by performing three kinds of conformal transformations in \cite{SRG}, where different accelerations have been found for accelerating quark and anti-quark pair. 
In this paper, based on the Nambu-Goto action we will consider an open string in $AdS_3$ with the following Poincare metric with $R=1$,
\begin{equation}
ds^2 = \frac{dz^2 - dt^2 + dx^2}{z^2}.
\label{me}\end{equation}
We make following ansatz \cite{JSH} for a rotating string with angular velocity $\omega$ which is moving with constant velocity $v$ in x direction from the AdS boundry at $z=0$ to the Poincare Horizon at $z=\infty$.
\begin{equation}
x=vt+r \sin{\omega t}
\label{xt}\end{equation}
In this work, the procedure of \cite{SRG} is followed to evaluate the effects of angular velocity on the accelerating quark and anti-quark pair. By performing the two mappings, the special conformal transformation and the conformal SO(2,2) transformation, we will obtain the acceleration of quark and anti-quark pair in six cases and we will see that quark and anti quark in some cases tend to deccelerate in order to approach each other then stop in accelerating in order to back away from each other.\\
So, we organize the paper as follows. In section 2 we discuss the case where we perform the special conformal transformations in four cases and we find the acceleration of quark and anti-quark and also we perform numerical calculation in order to investigate the effects of the angular velocity on the acceleration of the pair. In section 3 we perform the conformal SO(2,2) transformations as well. By the end, section 4 would be our conclusion.\\
\section{Special conformal transformations and rotating string}
As we know from \cite{SRG}, we can construct an open string configuration dual to accelerating quark and anti-quark by performing the special conformal transformations. In (\ref{ex}), $\frac{1}{b}$ is the acceleration of the quark and anti-quark pair \cite{BX} (see also \cite{SRG}).
Now, let us consider (\ref{xt}) where the string extends straight from the AdS boundary at $z=0$ to Poincare Horizon at $z=\infty$ in the $z$ direction and in addition to moving with constant velocity $v$ in the $x$ direction also it has rotation with angular velocity $\omega$.
If we consider $x=vt+r\sin{\omega t}$, we will face infinite loops in our numerical calculations, so we must use first order expression of extensions. On the other hand, our string profile becomes unstable then we take $\omega<<1$, which means that we are allowed to study very small fluctuations in this work. So our focuse is on the infenitesimal values of $\omega$, therefore we have,
\begin{equation}
\sin{\omega t}\approx\omega t.
\label{st}\end{equation}
Now, we intend to perform special conformal transformations of the Poincare $AdS_3$ spacetime coordinates $z, x^{\mu}=(t,x)$,
\begin{equation}
{x^{\mu}}' = \frac{x^{\mu} + a^{\mu}( z^2 + x^2 )}{1 + 2a\cdot x
+ a^2( z^2 + x^2 )}, \hspace{1cm} z' = \frac{z}{1 + 2a\cdot x
+ a^2( z^2 + x^2 )},
\label{pt}\end{equation}
the interested reader can see more details in \cite{BCF} that it has been studied for circular Wilson Loop.
As we know, we can construct particular $SL(2,R)_L$ and $SL(2,R)_R$ transformations by using special conditions. In order to obtain these particular transformations let us consider $SL(2,R)_L \times SL(2,R)_R$ isometry group which represents relations for $SL(2,R)_L$ and $SL(2,R)_R$ \cite{JMS, SR}. We can see both transformations map $AdS_3$ at $z=0$ to itself and act on the boundary as a usual conformal transformations of $1+1$ dimentional Minkowski space.\\
The $SL(2,R)_L$ transformation is,
\begin{eqnarray} 
w^+ \rightarrow {w^+}' &=& \frac{\alpha w^+ + \beta}{\gamma w^+ + \delta}, \nonumber \\
w^- \rightarrow {w^-}' &=& w^- + \frac{\gamma z^2}{\gamma w^+ + \delta}, \nonumber \\
z \rightarrow z' &=& \frac{z}{\gamma w^+ + \delta}
\end{eqnarray}
and $SL(2,R)_R$ transformation is
\begin{eqnarray} 
w^+ \rightarrow {w^+}' &=& w^+ + \frac{\gamma z^2}{\gamma w^- + \delta}, \nonumber \\
w^- \rightarrow {w^-}' &=& \frac{\alpha w^- + \beta}{\gamma w^- + \delta}, \nonumber \\
z \rightarrow z' &=& \frac{z}{\gamma w^- + \delta}. 
\end{eqnarray}
With real $\alpha, \beta, \gamma, \delta$ obeying $\alpha \delta - \beta \gamma = 1$.
Here we consider two special cases $a^{\mu} = (-\frac{1}{l},\frac{1}{l})$ and $a^{\mu} = (\frac{1}{l},\frac{1}{l})$ in (\ref{pt}) which cioncide with the particular $SL(2,R)_L$ and $SL(2,R)_R$ transformations with $\alpha = 1, \beta = 0, \gamma =2a$ and $\delta = 1$ respectively.
Let us make a special conformal transformation with $a^{\mu} = (-\frac{1}{l},\frac{1}{l})$ for the rotating string with angular velocity $\omega$ which is also moving with constant velocity $v$ in the $x$ direction
\begin{equation}
x=vt+r\omega t,
\label{om}\end{equation}
to have
\begin{eqnarray}
t' &=& \frac{1}{P}\left[t-\frac{1}{l}\left(z^2-t^2+v^2 t^2+r^2 \omega^2t^2+2vt^2r\omega\right)\right],\hspace{2cm}\nonumber \\
x' &=& \frac{1}{P}\left[vt+r\omega+\frac{1}{l}\left(z^2-t^2+v^2 t^2+r^2 \omega^2t^2+2vt^2r\omega\right)\right],\hspace{1cm}\nonumber \\
z' &=& \frac{z}{P},\hspace{6cm}
\label{op}\end{eqnarray}
where P is expressed in the following form
\begin{equation}
P = 1 + \frac{2}{l}\left(t+vt+r\omega t\right).
\label{rp}\end{equation}
By using $t'$, $x'$ in (\ref{op}) we obtain following expression for t 
\begin{equation}
t=\frac{P\left(x'+t'\right)}{1+v+r\omega},
\label{ft}\end{equation}
and with substitution of former expression in (\ref{rp}) we have P in terms of $t'$ and $x'$
\begin{equation}
P=\frac{l}{l-2t'-2x'}.
\end{equation}
Combining (2.6) and (2.9) leads to the following expanding string solution
\begin{equation}
\left(x'-\frac{l}{2\left(v+r\omega+1\right)}\right)^2=\left(t'-\frac{l\left(v+r\omega\right)}{2\left(v+r\omega+1\right)}\right)^2+\frac{l^2}{4}\frac{1-\left(v+r\omega\right)^2}{\left(v+r\omega+1\right)^2}-z'^2
\label{lr}\end{equation}
So we obtain an accelerating string configuration associated with proper acceleration $\frac{2\left(v+r\omega+1\right)}{l\sqrt{1-\left(v+r\omega\right)^2}}$ of a quark-antiquark pair.
Then we have found the term of acceleration of quark and antiquark pair by performing particular $SL(2,R)_L$ transformation. If we compare this expression with what it has been constructed in \cite{SRG} we will find that the angular velocity in this case acts like $v$. We perform some useful numerical calculations in order to show behaviors of the acceleration of the pair. We plot our results as Figure 1.\\
\begin{figure}
\centerline{\includegraphics[width=18cm]{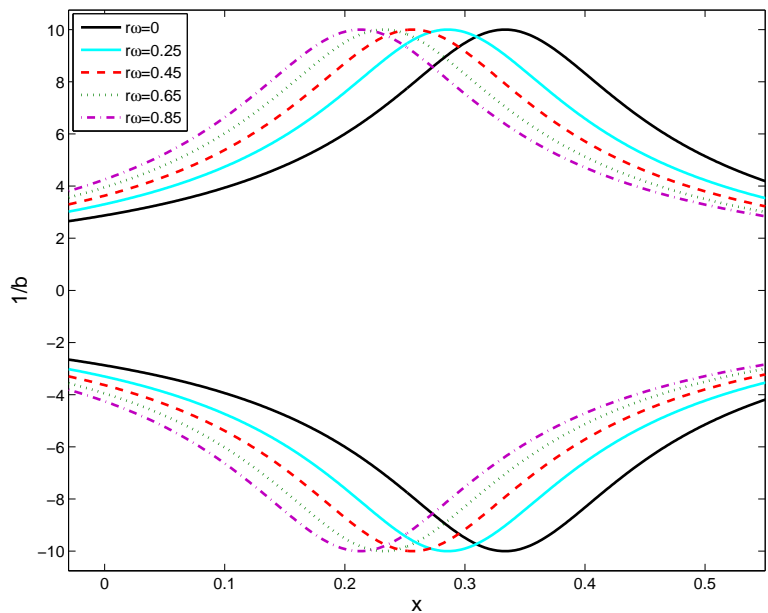}} \caption{This figure shows the behavior of acceleration of the pair, with respect to the $x$, where we have applied special conformal transformation with $a^{\mu}=(-\frac{1}{l},\frac{1}{l})$ which coincides with particular $SL(2,R)_L$ transformation. The solid black curve corresponds to $r\omega=0$, the solid blue curve corresponds to $r\omega=0.25$, the dashed red curve corresponds to $r\omega=0.45$, the dotted green curve coresponds to $r\omega=0.65$ and the purple dashed-dotted curve corresponds to $r\omega=0.85$.}
\end{figure}
In figure 1, we show the behavior of $\frac{1}{b}$ for this case as a function of $x$ for several choices of $r\omega$. In this figure and other figures in this paper, uper curves are ploted for plus sign and downer curves are ploted for minus sign, as one can see in (\ref{lr}) for this case. This figure shows that the maximum value of the acceleration is constant for different $r\omega$, it can be interpreted as permanent behavior of the accelerating quark and anti-quark pair with respect to any angular velocity in this transformation with $ a^{\mu}=(-\frac{1}{l},\frac{1}{l})$. So the pair has no feel to rotation but one can see that with increasing $r\omega$ curves are fined to be close to each other because quark and anti-quark get away from each other and it is not interesting for them. Also we see curves are shifted to left as a result of particular $SL(2,R)_L$ transformation. Shifting to the left means that existence of the maximum value of acceleration takes place in a shorter distance at more value of $\omega$.\\
Now we apply particular $SL(2,R)_R$ transformation by considering $a^{\mu}=(\frac{1}{l},\frac{1}{l})$, so we have,
\begin{eqnarray}
t' &=& \frac{1}{P}\left[t+\frac{1}{l}\left(z^2-t^2+v^2 t^2+r^2 \omega^2t^2+2vt^2r\omega\right)\right],\hspace{2cm}\nonumber \\
x' &=& \frac{1}{P}\left[vt+r\omega+\frac{1}{l}\left(z^2-t^2+v^2 t^2+r^2 \omega^2t^2+2vt^2r\omega\right)\right],\hspace{1cm}\nonumber \\
z' &=& \frac{z}{P},\hspace{6cm}
\label{sl}\end{eqnarray}
where P expressed in terms of $x'$ and $t'$ through
\begin{equation}
P=\frac{l}{l+2t'-2x'}.
\end{equation}
This expressions lead to following expanding string
\begin{equation}
\left(x'-\frac{l}{2\left(v+r\omega-1\right)}\right)^2=\left(t'-\frac{l\left(v+r\omega\right)}{2\left(v+r\omega-1\right)}\right)^2+\frac{l^2}{4}\frac{1-\left(v+r\omega\right)^2}{\left(v+r\omega-1\right)^2}-z'^2
\end{equation}
then the special conformal transformation with $a^{\mu}=(\frac{1}{l},\frac{1}{l})$ which coincides the particular $SL(2,R)_R$ transformation, generates the expanding string with acceleration $\frac{2(v+r\omega-1)}{l\sqrt{1-(v+r\omega)^2}}$.\\
\begin{figure}
\centerline{\includegraphics[width=18cm]{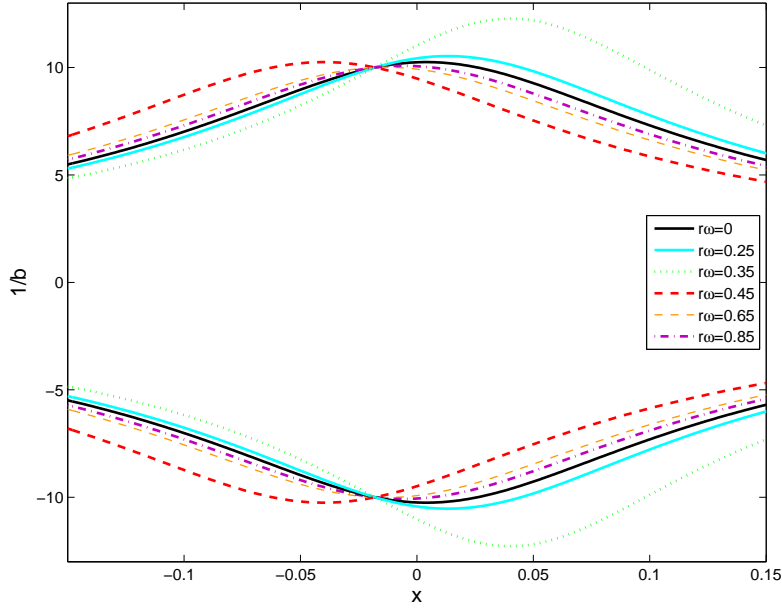}}
\caption{This figure shows the behavior of acceleration of the pair, with respect to the $x$, where we have applied special conformal transformation with $a^{\mu}=(\frac{1}{l},\frac{1}{l})$ which coincides with particular $SL(2,R)_R$ transformation. The solid black curve corresponds to $r\omega=0$, the solid blue curve corresponds to $r\omega=0.25$, the dotted light green curve corresponds to $r\omega=0.35$, the dashed red curve corresponds to $r\omega=0.45$, the dashed orange curve coresponds to $r\omega=0.65$ and the purple dashed-dotted curve corresponds to $r\omega=0.85$.}
\end{figure}
According to what we see in figure 2, we consider that when there is no angular velocity, which is studied in \cite{SRG}, maximum value of the acceleration happens in $x=0$, on the other hand there is no habitual relation between increasing $r\omega$ and maximum value of acceleration. The figure shows that in $r\omega=0.55$, unlike former curves, has a shift to left side and start decreasing. Then in this transformation quark and anti-quark aproach to each other in deccelerating and stop to return back in accelerating away from each other with proper acceleration. There is an interesting point in this figure, there is an exact position where the pair has constant value of acceleration, with any value for $r\omega$, it seems that in this position pair has no feel to $r\omega$ and just has its constant acceleration anyway.\\
We make another special conformal transformation with $a^{\mu} = (0,\frac{1}{l})$ for the rotating string moving with constant velocity v to have,
\begin{eqnarray}
t' &=& \frac{t}{P}, \nonumber \\
x' &=& \frac{1}{P}\left[ vt +r\omega t+\frac{1}{l}\left(z^2-t^2+v^2t^2+r^2\omega^2 t^2+2vt^2r\omega\right)\right], \nonumber \\
z' &=& \frac{z}{P},
\end{eqnarray}
We optain $x'$ in terms of $t'$ and $z'$ through
\begin{equation}
x' = \frac{l}{2} \pm \sqrt{\left( t' - \frac{l\left(v+r\omega\right)}{2}\right)^2 + 
\frac{(1-\left(v+r\omega\right)^2)l^2}{4} - {z'}^2 }
\end{equation}
So this special conformal transformation generates the expanding string with acceleration $\frac{2}{l\sqrt{1-(v+r\omega)^2}}$.
\begin{figure}
\centerline{\includegraphics[width=18cm]{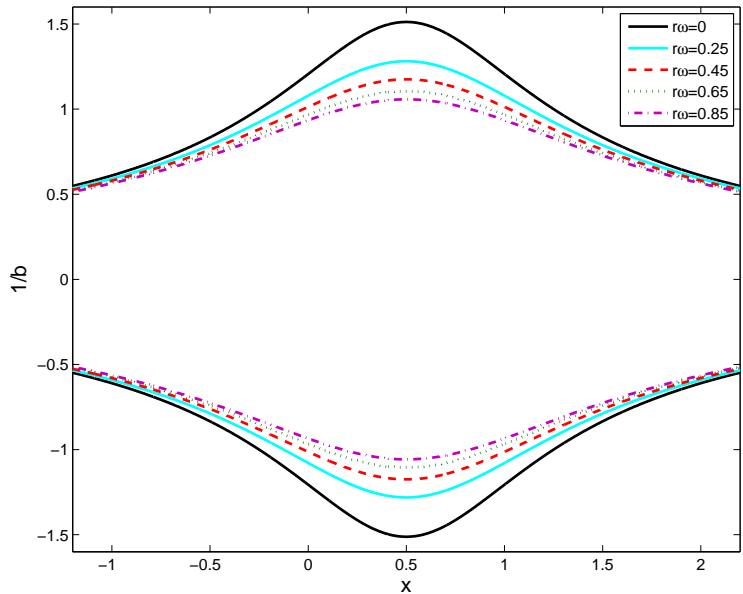}}
\caption{ This figure shows the behavior of acceleration of the pair, with respect to the $x$, where we have applied special conformal transformation with $a^{\mu}=(0,\frac{1}{l})$. The solid black curve corresponds to $r\omega=0$, the solid blue curve corresponds to $r\omega=0.25$, the dashed red curve corresponds to $r\omega=0.45$, the dotted green curve coresponds to $r\omega=0.65$ and the purple dashed-dotted curve corresponds to $r\omega=0.85$.}
\end{figure}
In this case we set time component in $a^{\mu}$ to zero, the Figure 3 shows that with increasing $r\omega$, acceleration reduces, so if $r\omega$ be large enough, pair has no acceleration any more. It means that quark and anti-quark aproach to each other in deccelerating and stop.
On the other hand, pair has acceleration in a finite range of $x$. thus, out of this region we consider curves are coincident, it means that the pair doesn't feel rotation.
On the other hand a special conformal transformation with $a^{\mu}=(\frac{1}{l},0)$ gives
\begin{equation}
x' = -\frac{l}{2\left(v+r\omega\right)} \pm \sqrt{\left( t' + \frac{l}{2}\right)^2 + 
\frac{(1-\left(v+r\omega\right)^2)l^2}{4\left(v+r\omega\right)^2} - {z'}^2 }.
\end{equation}
Which represents the expanding string with acceleration $\frac{2(v+r\omega/)}{l\sqrt{1-\left(v+r\omega\right)^2}}$.\\
\begin{figure}
\centerline{\includegraphics[width=18cm]{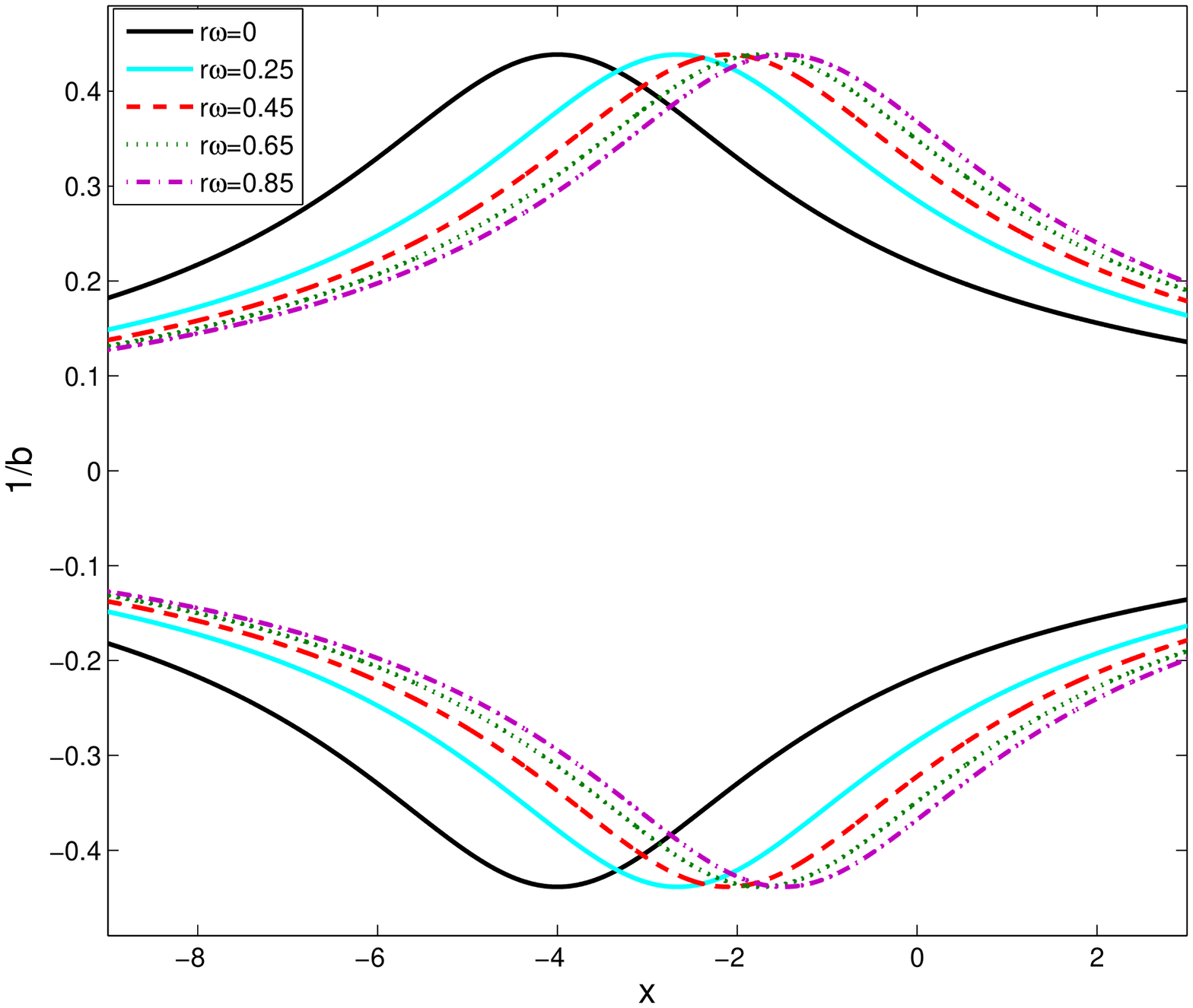}}
\caption{This figure shows the behavior of acceleration of the pair, with respect to the $x$, where we have applied special conformal transformation with $a^{\mu}=(\frac{1}{l},0)$. The solid black curve corresponds to $r\omega=0$, the solid blue curve corresponds to $r\omega=0.25$, the dashed red curve corresponds to $r\omega=0.45$, the dotted green curve coresponds to $r\omega=0.65$ and the purple dashed-dotted curve corresponds to $r\omega=0.85$.}
\end{figure}
In figure 4, unlike Figure 1, we see that increasing in angular velocity leads to shifts in positive $x$ direction (right side), also the maximum value of acceleration has no change. Shifting to the right side in larger values of $x$ shows that pair reaches to it's maximum value of acceleration in a larger distance $x$ with increasing $\omega$. It means that quark and antiquark move with constant acceleration.\\
As we show the accelerating quark and anti-quark pair which is constructed by former types of the special conformal transformations, can be affected by increasing in angular velocity, and they can aproach to each other or return back from each other in deccelerating or accelerating respectively.\\

\section{Conformal $SO(2,2)$ transformations and rotating string}
In this section we perform two conformal $SO(2,2)$ transformations on the rotating string which is also moving with constant velocity in the $x$ direction in order to study effects of angular velocity on the acceleration of quark and anti-quark.
Here we restore the AdS radius $R$ to express the following relations between the Poincare coordinates in $AdS_3$ and the embedding coordinates $X^M\; (M = -1, 0, 1, 2 )$ on which the conformal SO(2,2) transformation is acting linearly \cite{SRG},
\begin{eqnarray}
X^{\mu} &=& \frac{x^{\mu}}{z}R, \; (\mu =0, 1), \nonumber \\
X^{-1} &=& \frac{R^2 + z^2 + x_{\mu}x^{\mu} }{2z}, \hspace{1cm}
X^2 = \frac{R^2 - z^2 - x_{\mu}x^{\mu} }{2z}, \nonumber \\
- R^2 &=& - (X^{-1})^2 - (X^0)^2 + (X^1)^2 + (X^2)^2 .
\end{eqnarray}
For the string configuration (\ref{xt}), which is described
by $X^1 = (v+r\omega) X^0$, we perform following  conformal SO(2,2) transformation 
\begin{equation} 
{X^{-1}}' = - X^0, \;\; {X^0}' = X^{-1}, \;\; {X^1}' = X^1, \;\;
{X^2}' = X^2,
\label{os}\end{equation}
which interchanges $X^{-1}$ and $X^0$. The transformed configuration
is specified by ${X^1}' = - (v+r\omega) {X^{-1}}'$ which is expressed in terms of
the Poincare coordinates as
\begin{equation}
\left(x' + \frac{R}{v+r\omega}\right)^2 = {t'}^2 + 
\frac{1 - \left(v+r\omega\right)^2}{\left(v+r\omega\right)^2}R^2 - {z'}^2
\end{equation}
This expression represents the expanding string with acceleration $\frac{v}{R\sqrt{1-(v+r\omega)^2}}$.\\
\begin{figure}
\centerline{\includegraphics[width=18cm]{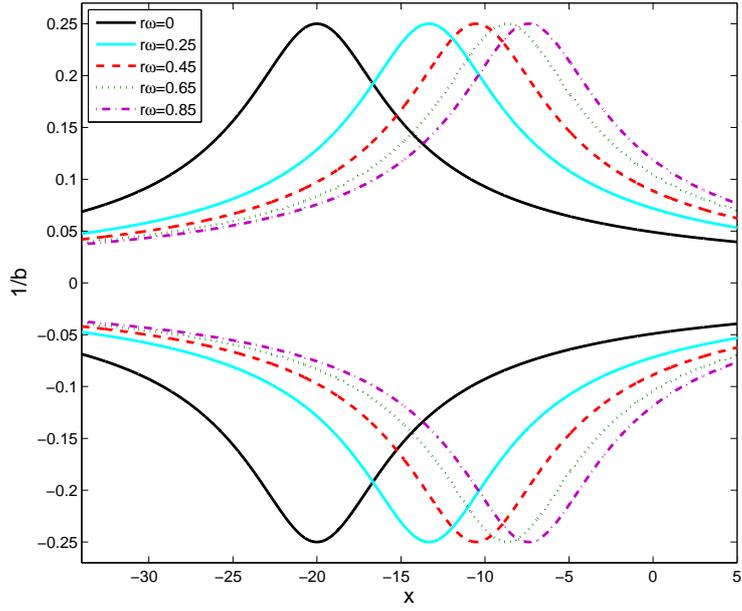}}
\caption{This figure shows the behavior of acceleration of the pair, with respect to the $x$, where we consider ${X^1}' = - (v+r\omega) {X^{-1}}'$. The solid black curve corresponds to $r\omega=0$, the solid blue curve corresponds to $r\omega=0.25$, the dashed red curve corresponds to $r\omega=0.45$, the dotted green curve coresponds to $r\omega=0.65$ and the purple dashed-dotted curve corresponds to $r\omega=0.85$.}
\end{figure}
In figure 5 we see the similar behavior with Figure 4, it shows that these transformations act similarly. As one can see, like the special conformal transformation with $a^{\mu}=(\frac{1}{l},0)$, this conformal $SO(2,2)$ transformation result in constant maximum value of acceleration in quark and anti-quark pair.\\
Now, let us apply the other conformal SO(2,2) transformation which has defined as the interchange 
between $X^1$ and $X^2$
\begin{equation}
{X^{-1}}' = X^{-1}, \;\; {X^0}' = X^{0}, \;\; {X^1}' = - X^2, \;\;
{X^2}' = X^1
\label{so}\end{equation}
So the transformed string configuration which is specified by  ${X^2}' = (v+r\omega) {X^0}'$, becomes\\
\begin{equation}
{x'}^2 = ( t' - \left(v+r\omega\right)R )^2 + (1 - (v+r\omega)^2)R^2 - {z'}^2
\end{equation}
thus we constructed the expanding string with acceleration $\frac{1}{R\sqrt{1-(v+r\omega)^2}}$.\\
\begin{figure}
\centerline{\includegraphics[width=18cm]{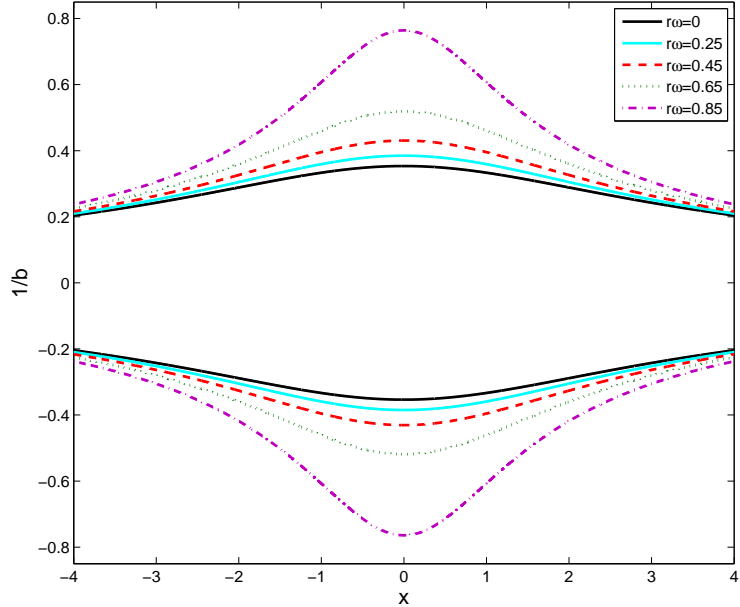}}
\caption{This figure shows the behavior of acceleration of the pair, with respect to the $x$, where we consider${X^2}' = (v+r\omega) {X^0}'$. The solid black curve corresponds to $r\omega=0$, the solid blue curve corresponds to $r\omega=0.25$, the dashed red curve corresponds to $r\omega=0.45$, the dotted green curve coresponds to $r\omega=0.65$ and the purple dashed-dotted curve corresponds to $r\omega=0.85$.}
\end{figure}
Figure 6 shows that this conformal transformation with ${X^2}' = (v+r\omega) {X^0}'$, unlike the special conformal transformation with $a^{\mu}=(0,\frac{1}{l})$, is the uniqe one in which maximum value of acceleration increases with increasing in $r\omega$, so maybe it is a chance for the quark and anti-quark pair to rotate with more values of $r\omega$ to achieve the large values of acceleration when it is sitting at $x=0$ position.\\

\section{Concolusion}
In this paper we have used the method of \cite{SRG} to investigate the effects of angular velocity on the accelerating quark and anti-quark in $AdS_3$ spacetime, and in general we have shown that increases in angular velocity can have effects on quark and anti-quark accelerating or deccelerating. We have considered and performed special conformal transformations and conformal SO(2,2) transformations in order to construct accelerating string configuration dual to accelerating quark and anti-quar pair. According to our numerical calculations and plots it is clarified that special conformal transformations whit $a^{\mu}=(-\frac{1}{l},\frac{1}{l})$ and $a^{\mu}=(\frac{1}{l},\frac{1}{l})$ which are coincide with $SL(2,R)_L$ and $SL(2,R)_R$ respectively, show behaviors which are not related or like other transformations. But we have shown that the special conformal transformation with $a^{\mu}=(0,\frac{1}{l})$ act contrariwise when angular velocity is increasing and it seems that if angular velocity becomes large enough deccelerating can happen, and conformal SO(2,2) transformation, where we consider ${X^2}' = (v+r\omega) {X^0}'$, vice versa. The special conformal transformation with $a^{\mu}=(\frac{1}{l},0)$ and the conformal SO(2,2) transformation, where we consider ${X^1}' = - (v+r\omega) {X^{-1}}'$, act similarly, as increases in angular velocity leads to constant maximum acceleration of quark and anti-quark pair. \\ 
\newpage 
\textbf{Acknowledgement}\\
We are grateful to Sara Tahery for useful discussions and also for comments on the manuscript.

\end{document}